\documentclass[superscriptaddress,twocolumn,showpacs,prb]{revtex4-1}
\usepackage[utf8]{inputenc} 
\usepackage{amsmath}
\usepackage{braket}
\usepackage{graphicx}
\usepackage{amsfonts}
\usepackage{pgfplots}
\usepackage{csquotes}
\usepackage{hhline}
\usepackage{amssymb}
\usepackage{listings}
\usepackage{color}

\definecolor{codegreen}{rgb}{0,0.6,0}
\definecolor{codegray}{rgb}{0.5,0.5,0.5}
\definecolor{codepurple}{rgb}{0.58,0,0.82}
\definecolor{backcolour}{rgb}{0.95,0.95,0.92}
 
\lstdefinestyle{mystyle}{
    backgroundcolor=\color{backcolour},   
    commentstyle=\color{codegreen},
    keywordstyle=\color{magenta},
    numberstyle=\tiny\color{codegray},
    stringstyle=\color{codepurple},
    basicstyle=\footnotesize,
    breakatwhitespace=false,         
    breaklines=true,                 
    captionpos=b,                    
    keepspaces=true,                 
    numbers=left,                    
    numbersep=5pt,                  
    showspaces=false,                
    showstringspaces=false,
    showtabs=false,                  
    tabsize=2
}
 
\usepackage{subfigure}

\usepackage{dcolumn}
\usepackage{tabularx}
\usepackage[colorlinks=true,linkcolor=blue,citecolor=blue,urlcolor=blue]{hyperref}
\usepackage{longtable}
\usepackage{braket}
\usepackage{float}
\newcolumntype{C}{>{\centering\arraybackslash}X}
\begin{document}
\title{Observation of Geometric Phase in a Molecular Aharonov-Bohm System Using IBM Quantum Computer}

\author{Gaurav Rudra Malik}
\email{gauravrudramalik@gmail.com}
\affiliation{Physics Department, \\Banaras Hindu University, Varanasi 221005, Uttar Pradesh, India}
\author{Sushree Swateeprajnya Behera }
\email{sushreeswateeprajnyabehera@gmail.com}
\affiliation{Department of Physics, \\ Indian Institute of Science Education and Research Tirupati 517507, Andhra Pradesh, India}

\author{Shubham Kumar}
\email{shubhamkumar.kumar@gmail.com}
\affiliation{Department of Physics,\\Central University of Jharkhand, Brambe 835205, Ranchi, India}

\author{Bikash K. Behera}
\email{bkb18rs025@iiserkol.ac.in}
\author{Prasanta K. Panigrahi}
\email{pprasanta@iiserkol.ac.in}
\affiliation{Department of Physical Sciences,\\ Indian Institute of Science Education and Research Kolkata, Mohanpur 741246, West Bengal, India}

\begin{abstract}

The evolution of a quantum system is governed by the associated Hamiltonian. A system defined by a parameter dependent Hamiltonian acquires a geometric phase when adiabatically evolved. Such an adiabatic evolution of a system having non-degenerate quantum states gives the well-studied Berry phase. Lounguet-Higgins and co-workers discovered a geometric phase when considering the Jahn-Teller distortion described by the nuclear coordinates traversing a closed path about the point of intersection of the electronic potential energy surfaces. Under such a condition, the Born-Oppenheimer wave function undergoes a sign change corresponding to an introduced global phase of $\pi$ radian. This change further introduces a multiple valuedness in the wavefunction which maybe removed by adding a vector potential like term in the Hamiltonian for the nuclear motion giving the Molecular Aharonov Bohm effect. Here, we demonstrate a scheme to evaluate the introduced global phase for molecular system considered by Longuet-Higgins and propose methods as first principle to do the same in more complex examples for the molecular Hamiltonian on a quantum computer.
\end{abstract} 

\maketitle
\section{Introduction}

Aharonov-Bohm effect shows that the wavefunction of a particle is affected by the presence of scalar and vector potentials, even in the absence of explicit electromagnetic fields\unskip~\cite{qbp_AharonovPR1959}. This effect is an instance of the Berry phase\unskip~\cite{qbp_BerryRoyalSociety1984} and is invariant under gauge transform as equivalent to the case of classical electrodynamics. This effect is due to the coupling of electromagnetic potential with the wave function of the charged particle. In the Born-Oppenheimer treatment of molecules, the electronic Hamiltonian and the electronic wave function depends on the nuclear coordinates, which acts as parameters\unskip~\cite{qbp_QuantumTheoryofMolecules}. The rapid evolution of the electronic wave function as compared to the nuclear displacement allows these wave functions to be considered separately. Since the time scales governing the electronic and nuclear progression differ widely, the nuclear displacement can be treated as the adiabatically varying parameter for electronic wavefunction\unskip~\cite{qbp_Mead_Review}. The adiabatic evolution of a qubit can be represented by the trajectory of the unit radial vector along the Bloch sphere. This equal to half the solid angle subtended by the traversed path at the origin\unskip~\cite{qbp_ErikS}. 

The molecular Hamiltonian has the kinetic and potential energy terms for the nuclear-electronic motion and coulombic interactions respectively. The inter electronic repulsion is studied using Hartree-Fock approximation. In addition the attractive electron-nuclear interaction term makes the Hamiltonian inseparable for electronic and nuclear parts; in the absence of the Born-Oppenheimer approximation. The electronic wave functions of the nuclear displacement must contain the nuclear coordinates as parameters which are adiabatically evolving, when considered independently. The variation of energy with respect to the nuclear coordinates of all possible configurations of the molecule gives a potential energy surface\unskip~\cite{qbp_PES_Hertzberg}, for instance, in a simple diatomic molecule; with respect to the inter-nuclear distance.

Hund proposed, in relation to the crossing of potential energy surfaces at the point of intersection, the electronic state must be degenerate. In polyatomic molecule, the intersection and the resulting degeneracy is a result of the variation of two or more internuclear distances and is hence a function of the molecular configuration. If two functions $\phi_1$ and $\phi_2$, are electronic wave functions corresponding to the intersecting potentials, then together with the functions for other energy levels, constitute an orthonormal set, as described by von Neumann and Wigner \unskip~\cite{qbp_PES_Hertzberg}. The electronic wave functions corresponding to the intersecting potentials is\unskip~\cite{qbp_PES_Hertzberg}:

\begin{equation}
\psi=c_1\phi_1 + c_2\phi_2
\label{qbp_Eq1}
\end{equation}

If $\hat{H}$ represents the electronic Hamiltonian in its matrix form then for degeneracy: $a_{11}= a_{22}$ and $a_{12}=0=a_{21}$, where $a_{ij}= \int \psi_i\hat{H}\psi_j^*$. This is analogous to the matrix equation obtained while considering perturbation in a two fold degenerate Hamiltonian. $\ a_{12} = 0$ requires the presence of two parameters in order to equate the real and imaginary terms equal to 0, and $\ a_{11}=a_{22}$ requires the presence of one parameter\unskip~\cite{qbp_PES_Teller}. For simplification, we may neglect the spin terms making the off diagonal terms real. For this, we require two parameters of molecular configuration for creating a degeneracy, which is the primary requirement for observing the crossing of potential surfaces.
The presence of two parameters makes it impossible for observing potential energy intersection in a diatomic molecule (except in special cases, e.g: Kramer's degeneracy) and the simplest example is the tri-atomic molecule with two parameters governing the molecular configuration. An outline following the Teller analysis of neglecting the spin terms of the electronic Hamiltonian was brought up by Herzberg and Longuet-Higgins in 1963\unskip~\cite{qbp_PES_Hertzberg}.

When two electronic degenerate states intersect at a point in real coordinate euclidean space, terms of the electronic Hamiltonian can be considered linearly dependent on the parameters, closer to the point of intersection. The Hamiltonian operator $\hat{H}$ is: 

\begin{equation}
\hat{H}=  \begin{bmatrix} 
\alpha_1x+\beta_1y & by \\
by & \alpha_2x+\beta_2y 
\end{bmatrix} 
\label{qbp_Eq2}
\end{equation}

Assuming the point of intersection at $x$=0 and $y$=0 and the corresponding energy $\epsilon$ = 0. For the degeneracy to hold, $\alpha_1$ = $\alpha_2$ and $\beta_1$ = $\beta_2$. For non zero energy $\epsilon$, we get the secular equation, which represents the equation of a double cone, with it's vertex at the point $\epsilon$=0. Hence the term `conical intersection'. This is given as\unskip~\cite{qbp_PES_Hertzberg,qbp_PES_Teller}:

\begin{equation}
    \epsilon = \pm[(\alpha x + \beta y)^2 + by]^\frac{1}{2}
    \label{qbp_Eq3}
\end{equation}

The electronic Hamiltonian operator $\hat{H}$ is: 

\begin{equation}
    \hat{H}=K \begin{bmatrix} 
1 & 0 \\
0 & 1 
\end{bmatrix}
    + K\begin{bmatrix} 
            \cos(\phi) & \sin(\phi) \\
            \sin(\phi) & -\cos(\phi) 
\end{bmatrix}
\label{qbp_Eq4}
\end{equation}
where K and $\phi$ are the parameterization of the coordinate space defined by substitutions $\alpha x = Ksin(\phi)$ and $\beta y =Kcos(\phi)$\unskip~\cite{qbp_PES_Hertzberg}. The eigenvalues of the operator are $ \pm $ K with the basis eigenvectors:

\begin{equation}
\begin{bmatrix} 
\cos(\phi/2)   \\
\sin(\phi/2) 
\end{bmatrix} and
\begin{bmatrix} 
-\sin(\phi/2)   \\
\cos(\phi/2) 
\end{bmatrix}
\label{qbp_Eq5}
\end{equation}

A variation of the parameter $\theta$ from $0$ to $2\pi$ makes the eigenstates multivalued, which is not allowed for a quantum state. To remove this multivaluedness, we include a global phase in the electronic wave function. If the multi-valued nature is removed, potential terms like the magnetic vector potential are introduced in the Born-Oppenheimer Hamiltonian\unskip~\cite{qbp_Mead_Review}.  
Let the electronic wave function be $\ket{\Psi(R)}$. It dependents on the nuclear coordinates $R$. After introducing the global phase we have\unskip~\cite{qbp_Mead_Review}:

\begin{equation}
\ket{\chi(R)}\longrightarrow\Ket{\chi(R)}e^{if(R)}  
\label{qbp_Eq6}
\end{equation}

According to Born Oppenheimer approximation, this wave function is separable and maybe written as: $\ket{\chi(R)}\psi(R)$. Here $\psi(R)$ is the nuclear wave function and $\Ket{\chi(R)}$ is the electronic eigenstate. This is hence the eigenstate of the $R$ dependent Born-Hamiltonian $\hat H(R)$. The Hamiltonian $\hat H(R)$ operates on the electronic eigenstate to give it's parameter dependent energy according to the eigenvalue equation $\hat H(R)\Ket{\chi(R)} = E(R)\Ket{\chi(R)}$. 

We define the nuclear momentum operator $\Hat P$ acting on the total wave function $\Psi(R)$ as an operator $\hat \Pi$ acting effectively on the nuclear wave function $\psi(R)$. This is represented as:

\begin{equation}
\hat\Pi \psi(R)=\left\langle \chi(R) \middle| \hat P \middle| \Psi (R) \right\rangle
\label{qbp_Eq7}
\end{equation}

This represents the inner product of the electronic wave function with the total wave function when operated upon by the momentum operator of the nuclear coordinates. The operator is represented by $\frac{1}{i}\nabla_R$. Differentiation is with respect to the nuclear coordinates. The above equation upon simplification yields:

\begin{equation}
\left\langle \chi(R) \middle| \frac{1}{i}\nabla_R \middle| \Psi (R) \right\rangle = \frac{1}{i}\nabla_R\psi(R) + A(R)\psi(R)
\label{qbp_Eq8}
\end{equation}

Here A(R) = $\frac{1}{i}\left\langle\chi(R)\middle| \nabla_R\chi(R) \right\rangle$. This is like the potential as magnetic vector potential associated with the momentum operator of the nuclear wave function, while considering the Born-Oppenheimer approximation\unskip~\cite{qbp_Mead_Review}. Considering the transformation in Eq. \eqref{qbp_Eq6}; the vector potential is:

\begin{equation}
A(R)\longrightarrow A(R) + \nabla f(R)   
\label{qbp_Eq9}
\end{equation}

This is the same as the gauge parameter when considering the gauge transform of the potentials leaving the electromagnetic fields invariant. Using the gauge transformation, the vector potential maybe reduced to zero implying $\left\langle\chi(R)\middle| \nabla_R\chi(R) \right\rangle$ =0 and hence $\left\langle\chi(R)\middle| \nabla_R\chi(R) \right\rangle\delta R$ = 0, representing the Berry parallel transport condition. The geometric phase in relation to parallel transport\unskip~\cite{qbp_Anandan}, is the evolution of a system where the phase remains unchanged with an infinitesimally small change in the state given that the norm of the state vector is constant throughout.

For every infinitesimal change in the nuclear coordinates $R$, the gauge term can be adjusted to make the vector potential term zero. However, the presence of the global phase term retains the multivalued nature of the eigenstates and we can either remove the multivalued nature of the wave function or make the vector potential vanish. In reference to Eq. \eqref{qbp_Eq5}; we may introduce a global phase factor of $\frac{\phi}{2}$ to make the eigenstates single-valued. As discussed above, a vector potential also accompanies the Hamiltonian, the magnitude of which maybe calculated using Eq. \eqref{qbp_Eq8}. 

Adiabatic transversal of nuclear coordinates is observed in the Jahn Teller distortion of molecules, describing the symmetry breaking within a molecular structure in order to attain a greater stability within. Octahedral complexes often shows the same behaviour, where the axial bonds are of different lengths compared to the equatorial bond lengths and some physical effects which includes magnetic properties of transition metal on compounds\unskip~\cite{qbp_Magnetism}. The symmetry breaking reduces the order of degeneracy and increases the stability\unskip~\cite{qbp_JahnTeller}. There are several simple theoretical models to study this phenomena which enable us to do numerical calculations\unskip~\cite{qbp_ModelJahnTeller}.

We evaluate the phase for eigenstates of Longuet-Higgins Hamiltonian, and propose a method for finding this phase for complex examples. We use the work by Viyuela \emph{et al.} for evaluating the topological Uhlmann Phase for the evolution of a mixed state\unskip~\cite{qbp_Delgado2018}. The Berry Phase has been estimated for a solid state qubit system about an adiabatically precessing magnetic field\unskip~\cite{qbp_WallraffScience2007}. Geometric phase has been observed in quantum gravitational cosmology theories accounting for the oscillating universe models\unskip~\cite{qbp_OscillatingCosmo} which might allow us to track back the formation of early universe according to big bang theory or the inflationary epoch\unskip~\cite{qbp_InflationCosmo}, its properties\unskip~\cite{qbp_Photonics_Majorana} are also used as a method for topological fault tolerant quantum computation\unskip~\cite{qbp_TopologicalQC}.    

\section{Methods}
In the simulation on the IBM simulator, we include two superconducting qubits; one for the system (S) and the other for the probe (P).

\begin{figure*}
\centering
\includegraphics[width=1\textwidth]{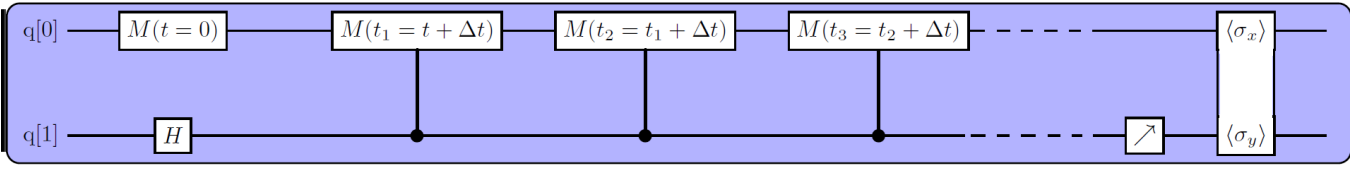} 
\caption{\textbf{General Circuit for evaluating phase in arbitrary number of steps}.}
\label{qbp_Fig1}
\end{figure*}

\textbf{Step 1}: Measuring the expectation values of $\sigma_x$ and $\sigma_y$ for the qubit P, fetches the real and imaginary parts of accumulated geometric phase, which is given by the inner product of the wave functions at the time $t$ and $t+\delta t$. The detailed protocol for measuring the above is described in the general circuit given in Fig. \ref{qbp_Fig1}.

\textbf{Step 2}: Applying the time evolution of wave function i.e., at $t_f=t+\delta t$, conditional to the state of the probe qubit, we get the combined state given by: 

\begin{equation}
\ket{\Psi}_{SP}=\frac{1}{\sqrt{2}}\Big(\Ket{\psi_{\theta(0)}}\otimes \Ket{0}_P+\Ket{\psi_{\theta(t_f)}}\otimes \Ket{1}_P\Big)
\label{qbp_Eq10}
\end{equation}

The probe qubit is in state $|1\rangle$, so $\Ket{\psi_{\theta(t_f)}}$ occurs as a tensor product with $\Ket{1_P}$. The retrieval of the geometric phase from the composite superposition of both the qubits involves taking the partial trace $\rho_P$ of the density matrix of the resultant composite system of S and P.

\textbf{Step 3}: From the state of the probe qubit after the holonomic evolution, we measure the geometric phase, by considering the reduced state of probe ($\rho_{_{P}}$). This is possible as the measurement is applied only on the probe qubit. By tracing out the system, we get in terms of the expectation values $\braket{\sigma_x}$ and $\braket{\sigma_y}$ the following expression for $\rho_{P}$. 

\begin{equation}
   \rho_{_P} = \frac{1}{2}\bigg(\mathbb I + Re [\left\langle \psi_{\theta(0)} \middle|  \psi_{\theta(t_f)} \right\rangle ]\sigma_x + Im[\left\langle \psi_{\theta(0)} \middle|  \psi_{\theta(t_f)} \right\rangle]\sigma_y\bigg)
   \label{qbp_Eq11}
\end{equation}
The geometric phase:
\begin{equation}
 \Phi = arg\big[\braket{\sigma_x}+i\braket{\sigma_y}\big]
 \label{qbp_Eq12}
\end{equation}
which are the expectation values of $\sigma_x$ and $\sigma_y$ on the output wave function.

\begin{figure}[]
\centering
\includegraphics[width=0.5\textwidth]{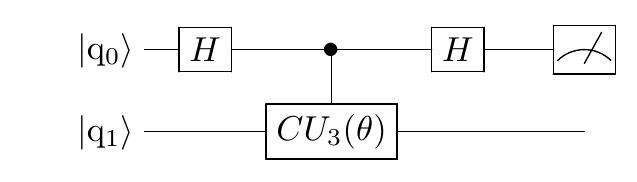}
\caption{\textbf{Quantum Circuit to evaluate phase in a single step}.}
\label{qbp_Fig2}
\end{figure}

\begin{figure}[]
\centering
\includegraphics[width=0.5\textwidth]{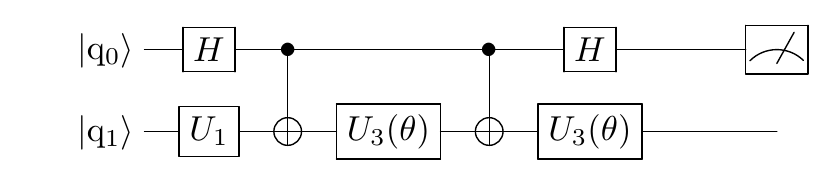}
\caption{\textbf{Corresponding Transpiled Circuit in IBM Q}.}
\label{qbp_Fig3}
\end{figure}

Evolution of the quantum state is given as an unitary evolution. We describe an operator which plays the role for the above of the state. The Hamiltonian is described by two parameters K and $\phi$ as described in Eq. \eqref{qbp_Eq4}. The phase factor arises due to the variation of the parameter $\phi$ from 0 to $2\pi$, as a function of time. Thus if the evolution is represented by the operator $M(t)$ then it's form is:

\begin{equation}
M(t) = e^{\int_{0}^{t}h(t')dt'}   
\label{qbp_Eq13}
\end{equation}
The expression for geometric phase can be modified as:
\begin{equation}
\Phi = arg\big[\braket{\psi_{_{\theta(0)}}|M(t_f)|\psi_{_{\theta(0)}} }\big]
\label{qbp_Eq14}
\end{equation}

The evolution operator represents the state of the initial wave function after time ``$t$". We assume that at time ``$t$" the angle changed is $\phi$. The initial state corresponds to superposition of the eigenstates of the Hamiltonian at a point when $\phi$ = 0.  

Using Eq. \eqref{qbp_Eq5}, the two eigenstates corresponding to $\phi$ = 0 are:
\begin{equation}
    \begin{bmatrix} 
      1  \\
      0 
\end{bmatrix}, 
     \begin{bmatrix} 
     0 \\
     1 
  \end{bmatrix}
  \label{qbp_Eq15}
\end{equation} 
which are the same as the basis state for a single qubit system. For an arbitrary angle $\phi$ we define the operation matrix $M(t)$ as:
\begin{equation}
M(t) = \sum|output\rangle \langle input|
\label{qbp_Eq16}
\end{equation}
Which implies,
\begin{equation}
   M =\begin{bmatrix} 
  \cos(\phi/2)   \\
            \sin(\phi/2)
            
   \end{bmatrix}
     \begin{bmatrix} 
    1 & 0\\
  \end{bmatrix}+
  \begin{bmatrix} 
  -\sin(\phi/2)   \\
            \cos(\phi/2)
  \end{bmatrix}
     \begin{bmatrix} 
    0 & 1\\
  \end{bmatrix}
  \label{qbp_Eq17}
\end{equation}
and hence we obtain the matrix,
\begin{equation}
   M(t)=\begin{bmatrix} 
  \cos(\phi/2)&-\sin(\phi/2)   \\
  \sin(\phi/2)& \cos(\phi/2)
  \end{bmatrix}
  \label{qbp_Eq18}
\end{equation}

The method described above by the Eq. \eqref{qbp_Eq16} is a straight forward method to calculate the operator matrix given the input and output states. Assuming orthogonal basis states as an input, the term remaining in the summation is the output corresponding to that particular input.

The state corresponding to the Hamiltonian parameter equal to $\phi$ at time $t$ is given by applying the above operator on the initial state. For preparing the initial state we apply the operator $M(\phi =0)$ to the basis states of the single qubit, assuming $\phi$ =0 at time $t=0$. This turns out to be the identity matrix.

The state of the system at time $t+\delta t$ is 
 \begin{equation}
M(t+\delta t) = e^{\int_{t}^{t+\delta t}h(t')dt'} 
\label{qbp_Eq19}
 \end{equation}

Assuming that the angle changed in time $\delta t$ is $\delta \phi$, the operator that evolves the wave function to the time $t+\delta t$ is:

\begin{equation}
   M(t+\delta t)=\begin{bmatrix} 
  \cos([\phi+\delta\phi]/2)&-\sin([\phi+\delta\phi]/2)   \\
  \sin([\phi+\delta\phi]/2)& \cos([\phi+\delta\phi]/2)
  \end{bmatrix}
  \label{qbp_Eq20}
\end{equation}

The application of this operator is conditional to the probe qubit. Since the eigenstates of the Hamiltonian given by Eq.\eqref{qbp_Eq5}, for $\phi$ = 2$\pi$ are given as,
\begin{equation}
    \begin{bmatrix} 
 -1  \\
 0
\end{bmatrix}, 
     \begin{bmatrix} 
    0 \\
    -1
  \end{bmatrix}
  \label{qbp_Eq21}
  \end{equation}

The matrix corresponding to this specific evolution may be obtained. As mentioned before, the eigenstates which are multivalued maybe represented by a transformation analogous to the Eq. \ref{qbp_Eq5} and hence be single valued. For this case the possible introduced phase term will be equal to $\pi$, as expected theoretically.

\section{Results}

The geometric phase is calculated by measuring the expectation value of the $\sigma_x$ and $\sigma_y$ operators of the output superposition state because it is not possible physically to measure the inner product between two states at separate time evolution on IBMq Experience platform. 

For $\sigma_x$, measure the output at a superposition of the eigenstates of the Pauli $X$ matrix i.e. the $\Ket{+}$ and $\Ket{-}$ state. We apply the Hadamard gate on the probe qubit just before the measurement to fascilate the process. In this case, the expectation value of $\sigma_x$ is given by the magnitude of ($P_o\Ket{0}-P_o\Ket{1}$), where $P_o$ is the notation for the probability of an event corresponding to the occurrence of the specific eigenstate either 0 or 1.

When $\phi$ initially is at $0$ and finally at $2\pi$, the quantum circuit becomes very straight forward and involves only 3 quantum gates in IBMq. The corresponding transpiled circuit which has been actually implemented on the machine is also shown along with the constructed circuit.

For the intermediate values of $\phi$, the circuit remains unchanged with the only varying parameter `theta' of the $U3$ gate. This evolution can be made to occur in a series of steps as described in the general circuit and can verify the parallel transport condition of evaluating the Berry phase.
The included graph shows the linear variation of the acquired geometric phase in steps of $\pi/6$ radian.

\begin{figure}[]
\centering
\includegraphics[width=0.5\textwidth]{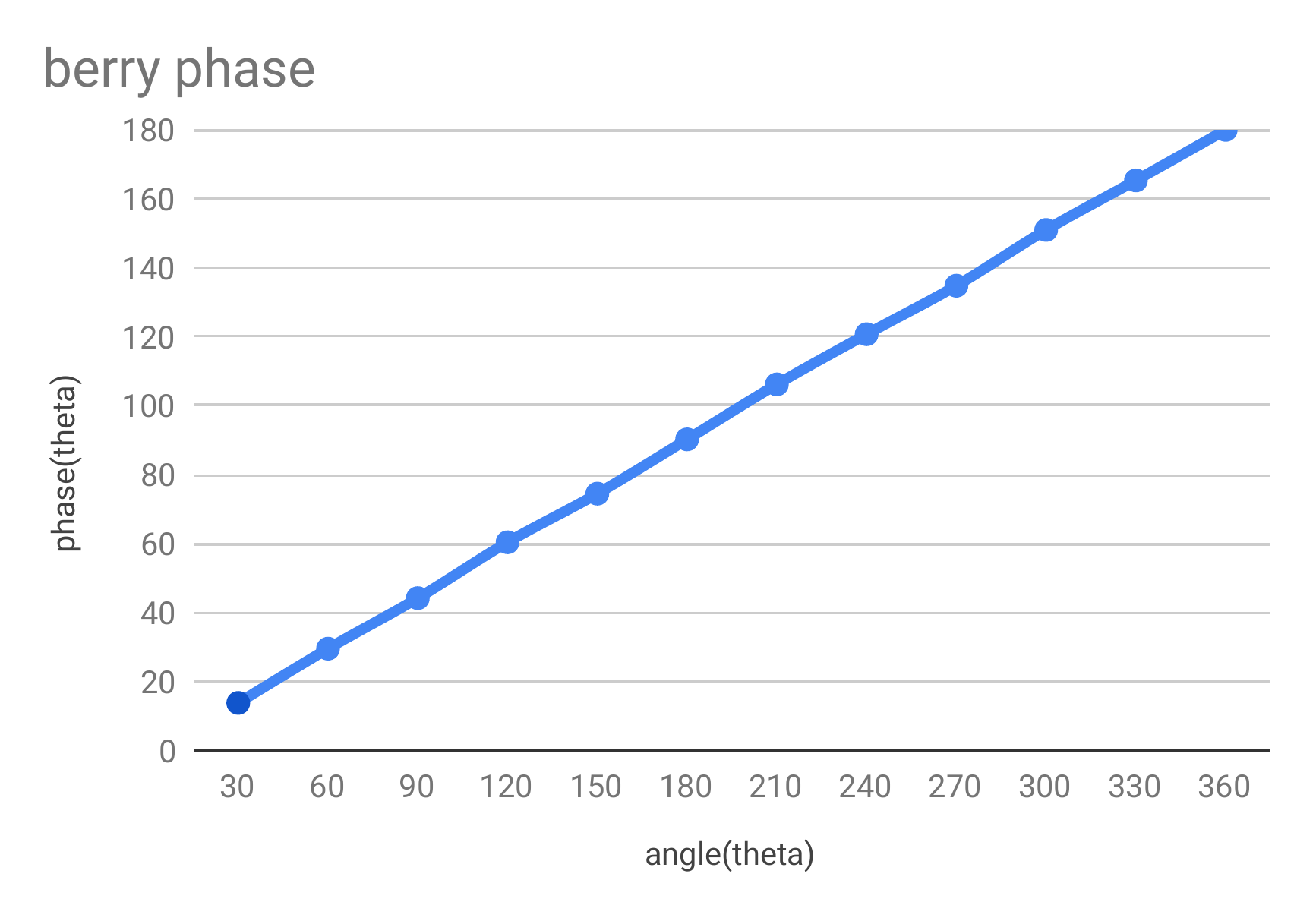}
\caption{\textbf{Accumulation of phase with rotation}.}
\label{qbp_Fig4}
\end{figure}

\begin{figure}[]
\centering
\includegraphics[width=0.5\textwidth]{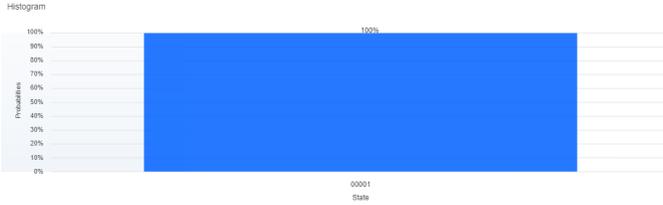}
\caption{\textbf{Simulation result of 2$\pi$ rotation} Implemented in Quantum Simulator.}
\label{qbp_Fig5}
\end{figure}

\begin{figure}[]
\centering
\includegraphics[width=0.5\textwidth]{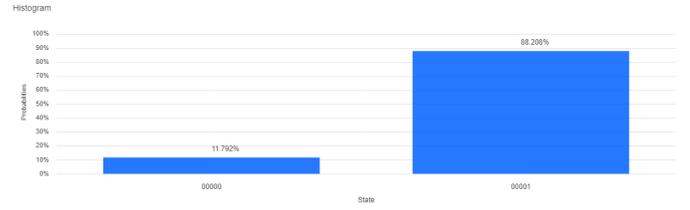}
\caption{\textbf{Simulation result of 2$\pi$ rotation}. Implemented in IBMqx4}
\label{qbp_Fig6}
\end{figure}

\begin{figure}[]
\centering
\includegraphics[width=0.5\textwidth]{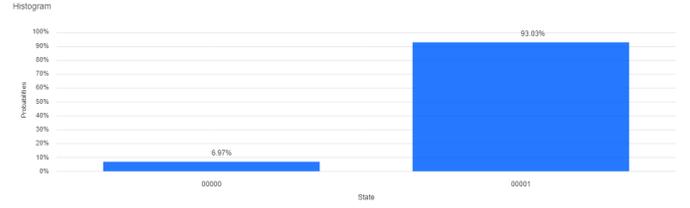}
\caption{\textbf{Simulation result of 2$\pi$ rotation}. Implemented in IBMqx2}
\label{qbp_Fig7}
\end{figure}

\section{Discussion}
Exact estimation of the geometric phase in context of the molecular Aharonov-Bohm system undergoing cyclic adiabatic evolution within the described protocol is reported. The experimental realization was done with the nominal circuit on the quantum simulator. This phenomena is useful in describing the Jahn-Teller effect, where a molecule undergoes distortion thereby lowering its symmetry in order to be more stable. We have considered a simple example, considered by Longuet-Higgins to describe this phenomena in terms of adiabatically varying nuclear coordinates\unskip~\cite{qbp_Berry_Nature}. 
For a complex molecule having larger number of inter nuclear parameters, the order of degeneracy introduced is higher. The corrected perturbed wave function maybe represented as a linear combination of the unperturbed wave functions, therefore we have a secular determinant of the order $m*m$, where $m$ represents the order of degeneracy. Solving the secular determinant gives $m$ roots corresponding to the levels into which the degenerate energy level has split.
 
The $e\otimes E$ Jahn-Teller effect arises from the vibronic coupling, for which the electron-nuclear term in the molecular Hamiltonian can be expressed in terms of normal mode coordinates $Q_i$ of the molecule. The vibronic coupling terms exist as perturbations as:
 
\begin{equation}
\hat{H}= H_0 + \Sigma_1 \hat{V_1}Q_1 + \frac{1}{2}\Sigma_{1,2}\hat{V_{1,2}}Q_{1,2} + ...   
\label{qbp_Eq22}
\end{equation}
 
Here $\hat{V_1} = \frac{\partial H}{\partial Q_1}$ and $\hat{V_{1,2}} = \frac{\partial^2 H}{\partial Q_1\partial Q_2}$ and are the first and second order perturbations respectively. $\hat{H_0}$ is the electronic Hamiltonian at the equilibrium nuclear configuration. 
 
Let \{$\Phi_i^{(0)}$\} form an orthonormal set of the unperturbed electronic equilibrium Hamiltonian. The secular matrix is :
  
\begin{equation}
\sum_{j=1}^{m} \left\langle \Phi_i^{(0)} \middle| \hat{V} \middle| \Phi_j^{(0)} \right\rangle - E_n^{(1)} \left\langle \Phi_i^{(0)} | \Phi_j^{(0)} \right\rangle 
\label{qbp_Eq23}
\end{equation}
 
$\therefore V_{ij} = \sum_{k} Q_{k}\left\langle \Phi_i^{(0)} \middle|\hat{V_k}  \middle| \Phi_j^{(0)} \right\rangle$ represents the Hamiltonian accounting first order perturbation terms. When the normal modes are considered as the equilibrium nuclear positions (\{$Q_k$\}=0 $\equiv Q^0$), above terms vanish. If at $Q^0$ the degenerate wave functions $\Phi_i^(0)$ transform according to the irreducible representation $\Gamma^\alpha$ along with $V_k$ and $Q_k$ transforming according to $\Gamma^\beta$, the totally symmetric transform ($\Gamma^0$) must be a subset of $\Gamma^\beta\otimes[\Gamma^\alpha]^2$ for Jahn-Teller effect to take place. Here $[\Gamma^\alpha]^2$ is the symmetrized product of $\Gamma^\alpha$.
 
Under such specific conditions the evaluation of the geometric phase under adiabatic evolution of the Hamiltonian, becomes numerically very intensive. There are hybrid quantum machine learning algorithms which makes it an efficient process on obtaining the eigenstates of the most complex operators\unskip~\cite{qbp_HybridEigen}. We employ the quantum simulator and the controlled coupling of the two qubits for this process by the means of a state-independent protocol to identify the exact nature of the introduced geometric phase.

\section*{Acknowledgments}
\label{qlock_acknowledgments}
G.R.M., S.S.B. and S.K. would like to thank IISER Kolkata for providing hospitality during the course of the project. B.K.B. acknowledges the support of Institute fellowship provided by IISER Kolkata. G.R.M and S.K. acknowledge support of the QIQT Fellowship. The authors acknowledge the support of IBM Quantum Experience. The views expressed are those of the authors and do not reflect the official policy or position of IBM or the IBMq Experience team.

\clearpage


\begin{thebibliography}{10}
\expandafter\ifx\csname url\endcsname\relax
  \def\url#1{\texttt{#1}}\fi
\expandafter\ifx\csname urlprefix\endcsname\relax\def\urlprefix{URL }\fi
\providecommand{\bibinfo}[2]{#2}
\providecommand{\eprint}[2][]{\url{#2}}

\bibitem{qbp_AharonovPR1959}
\bibinfo{author}{Y. Aharonov, D. Bohm}
\newblock \bibinfo{title}{Significance of Electromagnetic Potentials in the Quantum Theory}.
\newblock \emph{\bibinfo{journal}{The Physical Review}} \textbf{\bibinfo{volume}{115}},
  \bibinfo{pages}{485} (\bibinfo{year}{1959}).

\bibitem{qbp_BerryRoyalSociety1984}
\bibinfo{author}{M.V Berry}   
\newblock \bibinfo{title}{Quantal phase factors accompanying adiabatic changes}
\newblock \emph{\bibinfo{journal}{Proceedings of the Royal Society A}} \textbf{\bibinfo{volume}{392}}, \bibinfo{pages}{1802} (\bibinfo{year}{1984}).
 
\bibitem{qbp_QuantumTheoryofMolecules}
\bibinfo{author}{Brian T. Sutcliffe, R. Guy Woolley}   
\newblock \bibinfo{title}{Quantum Theory of Molecules}
\newblock \emph{\bibinfo{arxiv preprint}{arXiv:1206.4239 [quant-ph]}} 

\bibitem{qbp_Mead_Review}
\bibinfo{author}{C. Allen Mead}   
\newblock \bibinfo{title}{The geometric phase in Molecular Systems}
\newblock \emph{\bibinfo{journal}{Review Modern Physics}} \textbf{\bibinfo{volume}{64}}, \bibinfo{pages}{51} (\bibinfo{year}{1992})

\bibitem{qbp_ErikS}
\bibinfo{author}{Eric Sj$\Ddot{o}$qvist}   
\newblock \bibinfo{title}{}
\newblock \emph{\bibinfo{journal}{arxiv pre print arXiv:1503.04847v2 [quant-ph] }} (\bibinfo{year}{2015})

\bibitem{qbp_PES_Hertzberg}
\bibinfo{author}{G. Hertzburg and H.C. Longuet-Higgins}   
\newblock \bibinfo{title}{Intersection of potential energy surfaces in polyatomic molecules}
\newblock \emph{\bibinfo{journal}{Discussions of the Faraday Society}} \textbf{\bibinfo{volume}{35}},(\bibinfo{year}{1963})

\bibitem{qbp_PES_Teller}
\bibinfo{author}{E. Teller}   
\newblock \bibinfo{title}{Crossing of potential energy surfaces}
\newblock \emph{\bibinfo{journal}{American Chemical Society}},(\bibinfo{year}{1936})

\bibitem{qbp_Anandan}
\bibinfo{author}{J. Anandan and Y.Aharonov}   
\newblock \bibinfo{title}{Geometric Quantum Phase and Angles}
\newblock \emph{\bibinfo{journal}{Physics Review D}},\textbf{\bibinfo{volume}{38}}, \bibinfo{pages}{1863} (\bibinfo{year}{1988})

\bibitem{qbp_Magnetism}
\bibinfo{author}{Kliment I Kugel' and D I Khomskiĭ}   
\newblock \bibinfo{title}{The Jahn-Teller effect and magnetism: transition metal compounds}
\newblock \emph{\bibinfo{journal}{Soviet Physics Upekhi}},\textbf{\bibinfo{volume}{25}}, (\bibinfo{year}{1982})

\bibitem{qbp_JahnTeller}
\bibinfo{author}{Mary C.M.O'Brien and C.C.Chancey}   
\newblock \bibinfo{title}{The Jahn-Teller effect: An introduction and Current Review}
\newblock \emph{\bibinfo{journal}{American Journal of Physics}},\textbf{\bibinfo{volume}{61}}, \bibinfo{pages}{688} (\bibinfo{year}{1993})

\bibitem{qbp_ModelJahnTeller}
\bibinfo{author}{Peter Senn}   
\newblock \bibinfo{title}{A simple quantum mechanical model that illustrates the Jahn-Teller effect}
\newblock \emph{\bibinfo{journal}{Journal of Chemical Education}},\textbf{\bibinfo{volume}{69}}, \bibinfo{pages}{819} (\bibinfo{year}{1992})


\bibitem{qbp_Delgado2018}
\bibinfo{author}{O. Viyuela,, A. Rivas, S. Gasparinetti, A. Wallraff, S. Filipp, M. A. Martin-Delgado.} \bibinfo{author}{Rivas, A.} \bibinfo{author}{Martínez, E.~ A.}\bibinfo{author}{Nigg, D.}\bibinfo{author}{Schindler, P.}\bibinfo{author}{Monz, T.}\bibinfo{author}{Blatt, R.}\bibinfo{author}{Martin-Delgado, M.~ A.}

\bibitem{qbp_WallraffScience2007}
\bibinfo{author}{P. J. Leek, J. M. Fink, A. Blais, R. Bianchetti, M. Göppl, J. M. Gambetta,, D. I. Schuster, L. Frunzio, R. J. Schoelkopf, A. Wallraff} 
\newblock \bibinfo{title}{Observation of Berry's Phase in a Solid-State Qubit}.
\newblock \emph{\bibinfo{journal}{{Science}}\textbf{\bibinfo{volume}{318}},
\bibinfo{pages}{1889-1892} (\bibinfo{year}{2007})}.

\bibitem{qbp_InflationCosmo}
\bibinfo{author}{Barun Kumar Pal, Supratik Pal, B. Basu}   
\newblock \bibinfo{title}{Berry's Phase in Inflation Cosmology}
\newblock \emph{\bibinfo{journal}{arxiv pre print arXiv:1108.3689 [astro-ph.CO}}, (\bibinfo{year}{2013})

\bibitem{qbp_OscillatingCosmo}
\bibinfo{author}{Paul J Steinhardt and Neil Turok}   
\newblock \bibinfo{title}{A Cyclic Model of the Universe}
\newblock \emph{\bibinfo{journal}{Physics Review D}},\textbf{\bibinfo{volume}{296}}, \bibinfo{pages}{1436} (\bibinfo{year}{2002})

\bibitem{qbp_Photonics_Majorana}
\bibinfo{author}{ Jin-Shi Xu, Kai Sun, Jiannis K. Pachos, Yong-Jian Han, Chuan-Feng Li and Guang-Can Guo}   
\newblock \bibinfo{title}{Photonic implementation of Majorana-based Berry phases}
\newblock \emph{\bibinfo{journal}{Science Advances}} \textbf{\bibinfo{volume}{4}}, \bibinfo{pages}{10} (\bibinfo{year}{2018})

\bibitem{qbp_TopologicalQC}
\bibinfo{author}{Michael H. Freedman, Alexei Kitaev, Michael J. Larsen, Zhenghan Wang}   
\newblock \bibinfo{title}{Topological Quantum Computation}
\newblock \emph{\bibinfo{journal}{arxiv pre print: arXiv:quant-ph/0101025}} (\bibinfo{year}{2002})

\bibitem{qbp_Berry_Nature}
\bibinfo{author}{M.V Berry}   
\newblock \bibinfo{title}{Geometric Phase Memories}
\newblock \emph{\bibinfo{journal}{Nature Physics}} \textbf{\bibinfo{volume}{6}}, \bibinfo{pages}{148} (\bibinfo{year}{2010})

\bibitem{qbp_HybridEigen}
\bibinfo{author}{F. Albarrán-Arriagada, J. C. Retamal, E. Solano, L. Lamata}   
\newblock \bibinfo{title}{Reinforcement learning for semi-autonomous approximate quantum eigensolver}
\newblock \emph{\bibinfo{journal}{arxiv pre print: arXiv:1906.06702 [quant-ph]}} (\bibinfo{year}{2019})



\end{thebibliography}
\end{document}